\documentclass[aps,prl,reprint,showpacs,superscriptaddress,nofootinbib]{revtex4-1}

\usepackage{graphicx}
\usepackage{amsmath}
\usepackage{amsfonts}

\def\vep{\varepsilon}
\DeclareMathOperator{\Tr}{Tr}
\DeclareMathOperator{\rea}{Re}
\DeclareMathOperator{\ima}{Im}

\begin{document}

\title{Laughlin spin liquid states on lattices obtained from conformal field theory}
\author{Anne E. B. Nielsen}
\author{J. Ignacio Cirac}
\affiliation{Max-Planck-Institut f{\"u}r Quantenoptik,
Hans-Kopfermann-Strasse 1, D-85748 Garching, Germany}
\author{Germ\'an Sierra}
\affiliation{Instituto de F\'isica Te\'orica, UAM-CSIC, Madrid, Spain}

\begin{abstract}
We propose a set of spin system wavefunctions that are very similar to lattice versions of the Laughlin states. The wavefunctions are conformal blocks of conformal field theories, and for filling factor $\nu=1/2$ we provide a parent Hamiltonian, which is valid for any even number of spins and is at the same time a 2D generalization of the Haldane-Shastry model. We also demonstrate that the Kalmeyer-Laughlin state is reproduced as a particular case of this model. Finally, we discuss various properties of the spin states and point out several analogies to known results for the Laughlin states.
\end{abstract}


\maketitle

Phenomena in strongly correlated systems are generally hard to understand and describe, and therefore simple model systems exhibiting various behaviors are important guides. Laughlin's wavefunctions \cite{laughlin}
\begin{equation}\label{Laughlin}
\prod_n\tilde{\chi}(Z_n)\prod_{n<p}(Z_n-Z_p)^{1/\nu}\exp\left(-\sum_q|Z_q|^2/4\right),
\end{equation}
e.g., have played a key role in explaining the fractional quantum Hall effect, and this has triggered interest in understanding the nature of these states in depth. (The complex numbers $Z_n$ are the positions of the particles in the complex plane, $\nu$ is the Landau level filling factor, and the product of the single-particle phase factors $\tilde{\chi}(Z_n)$ is the gauge factor.) Finding parent Hamiltonians for wavefunctions is also very useful because it tells us how the behavior described by the wavefunction can be associated to the interactions between the particles in the system, and it can guide us to find experimental situations where such a behavior occurs, even if the Hamiltonian itself cannot be implemented directly.

In 1987, Kalmeyer and Laughlin (KL) introduced the $\nu=1/2$ bosonic Laughlin state on a square lattice \cite{laughlinstate}. Such models are expected to play a crucial role in understanding topological phases of lattice systems, in very much the same way as Laughlin states do in bulk. Furthermore, they may open up the door \cite{sorensen} for the experimental realization and investigation of Laughlin-like states under very well-controlled conditions in optical lattices. The KL state has been further investigated in \cite{laughlinHam,schroeter}, but it has been a long-standing problem to find a parent Hamiltonian for the state. Within the last few years, Hamiltonians have been found that are exact in the thermodynamic limit \cite{greiter,thomale,kapit}. In the present paper, we take a different approach, in which we propose to slightly modify the Laughlin states. The modification enables us to find a relatively simple Hamiltonian, containing only two- and three-body interactions, which is exact also for finite systems and arbitrary lattices. We demonstrate that the modified states are very close to the original Laughlin lattice states for finite systems, while they are exactly the same in the limit of an infinite square lattice.

As in the spirit of \cite{MR}, the wavefunctions we propose are chiral correlators of conformal blocks. The key element in the derivation of the Hamiltonian is to exploit that this structure allows us to apply rules from conformal field theory (CFT). Using the properties of null fields in CFT, we have recently derived \cite{nsc} nonuniform and higher spin generalizations of the 1D Haldane-Shastry (HS) model \cite{H88,S88}, and in the present paper, we extend these results further to obtain a generalization of the HS model to 2D. Previous work on finding parent Hamiltonians for the KL state on an infinite lattice has in part been inspired by the original HS model. Here, we complete this idea by demonstrating that the HS state and the KL state are in fact two limiting cases of the same model.

In addition, we characterize the most important physical properties of the modified states, including correlation functions, topological entanglement entropies, and entanglement spectra. It has been noted numerically that for some fractional quantum Hall states, the entanglement spectrum corresponds to the spectrum of the CFT that defines the wavefunction on the boundary \cite{haldane}. Utilizing the particular structure of the proposed wavefunctions, we are here able to show analytically that the entanglement spectrum for a two-legged ladder (as defined in \cite{poilblanc}) exactly corresponds to the 1D CFT with central charge $c=1$ for $\nu=1/2$, and we trace this back to the fact that the Yangian symmetry is inherited at the boundary.

\paragraph{Wavefunction}

The wavefunctions we propose describe the state of $N$ spin $1/2$ particles at positions $z_1,\ldots,z_N$ in the complex plane, where $N$ is even. They are chiral correlators of products of vertex operators $\phi_{s_n}(z_n)={:e^{i\sqrt{\alpha}s_n\varphi(z_n)}:}$ \cite{cft-book}, where $:\ldots:$ means normal ordering, $\alpha$ is a positive parameter, $s_n=\pm1$ is twice the $z$-component of the $n$'th spin, and $\varphi(z_n)$ is the field of a free mass-less boson, i.e.\ \cite{cft-book},
\begin{multline}\label{wf}
\psi_{s_1,\ldots,s_N}(z_1,\ldots,z_n)
=\langle \phi_{s_1}(z_1)\cdots\phi_{s_N}(z_N)\rangle\\
=\delta_\mathbf{s}\prod_{p=1}^N\chi_{p,s_p}\prod_{n<m}^N(z_n-z_m)^{\alpha s_ns_m}.
\end{multline}
Here, $\delta_\mathbf{s}=1$ for $\sum_ns_n=0$ and $\delta_\mathbf{s}=0$ otherwise, and the phase factors $\chi_{p,s_p}$ can be chosen at will, since the chiral correlator is only defined up to a phase. For $\alpha=1/2$, we shall always choose $\chi_{p,s_p}=\exp[i\pi(p-1)(s_p+1)/2]$, since this ensures that \eqref{wf} is a singlet \cite{supplement}.

For $\alpha=1/4$, we note that \eqref{wf} can be written (up to an overall phase) as a Slater determinant of the single-particle wavefunctions $\psi_k(z_n)=z_n^{k-1}(\chi_{n,1}/\chi_{n,-1})\prod_{m(\neq n)}(z_n-z_m)^{-1/2}$, $k=1,2,\ldots,N/2$. (We use the convention that $\prod_{n\neq m}$ is the product over $m$ and $n$, whereas $\prod_{m({\neq}n)}$ is the product only over $m$.) The states $\psi_k(z_n)$ are not orthonormal, but can be made so without changing the Slater determinant. We can therefore regard \eqref{wf} as the state of $N/2$ noninteracting fermions. This simplification enables us to use exact numerical computations rather than Monte Carlo simulations for $\alpha=1/4$ when we compute properties of the wavefunctions below. The $\nu=1$ Laughlin state can also be written as a Slater determinant, and indeed we shall see in a moment that $\alpha=1/4$ corresponds to $\nu=1$.

\paragraph{Connection to the Laughlin states}

We next investigate the statement that \eqref{wf} is similar to lattice versions of the Laughlin states, which are in turn closely related to the continuous Laughlin states. We expect the correspondence to be approximately valid for all lattice configurations for which the distribution of the lattice points is not too far from uniform and also if the complex plane is mapped into other geometries. We shall here consider the case of a (finite) square lattice in the complex plane since it is mathematically convenient and the case of an approximately uniform distribution on the sphere because this geometry eliminates all boundaries.

\begin{figure}
\includegraphics[width=0.49\columnwidth]{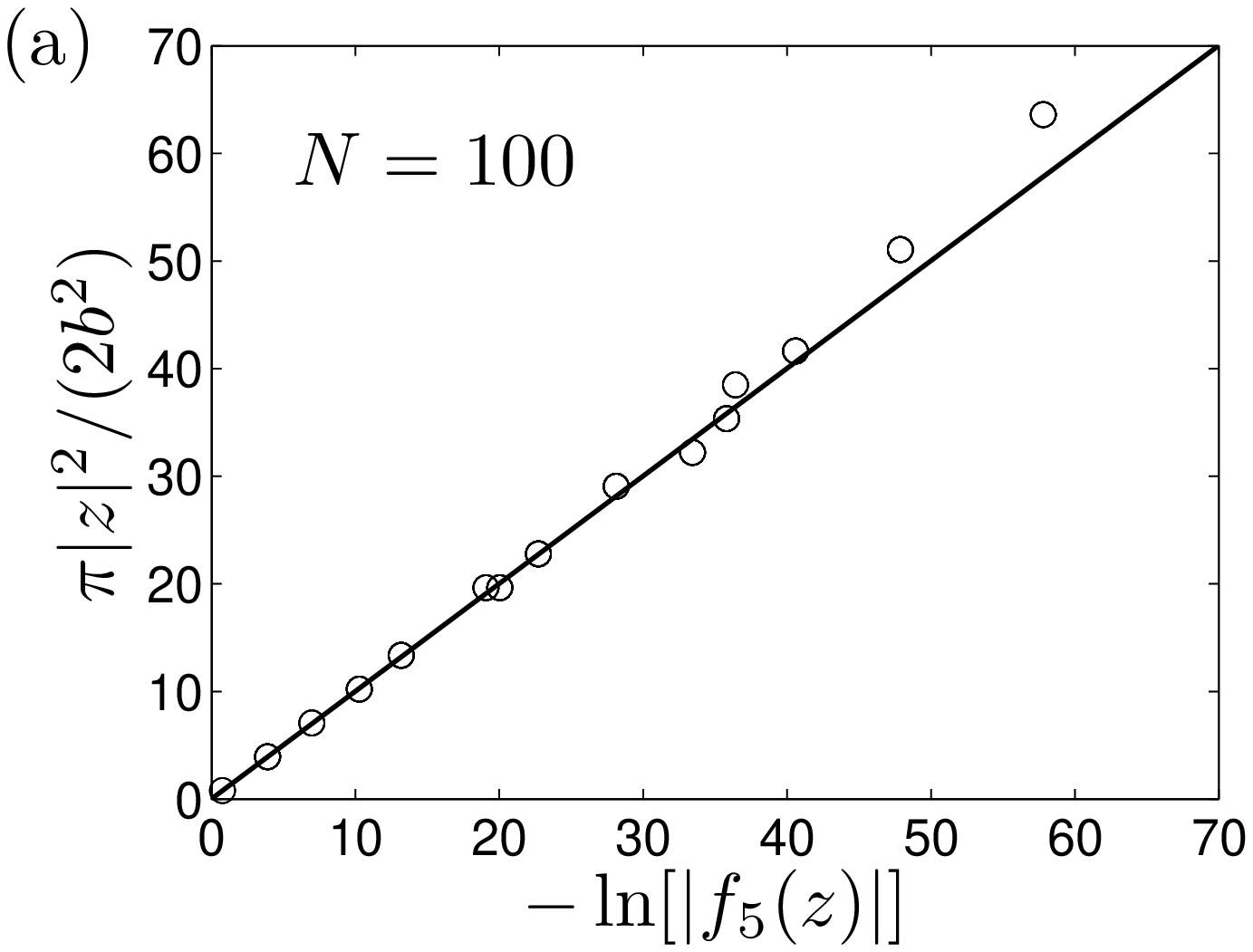}
\includegraphics[width=0.49\columnwidth]{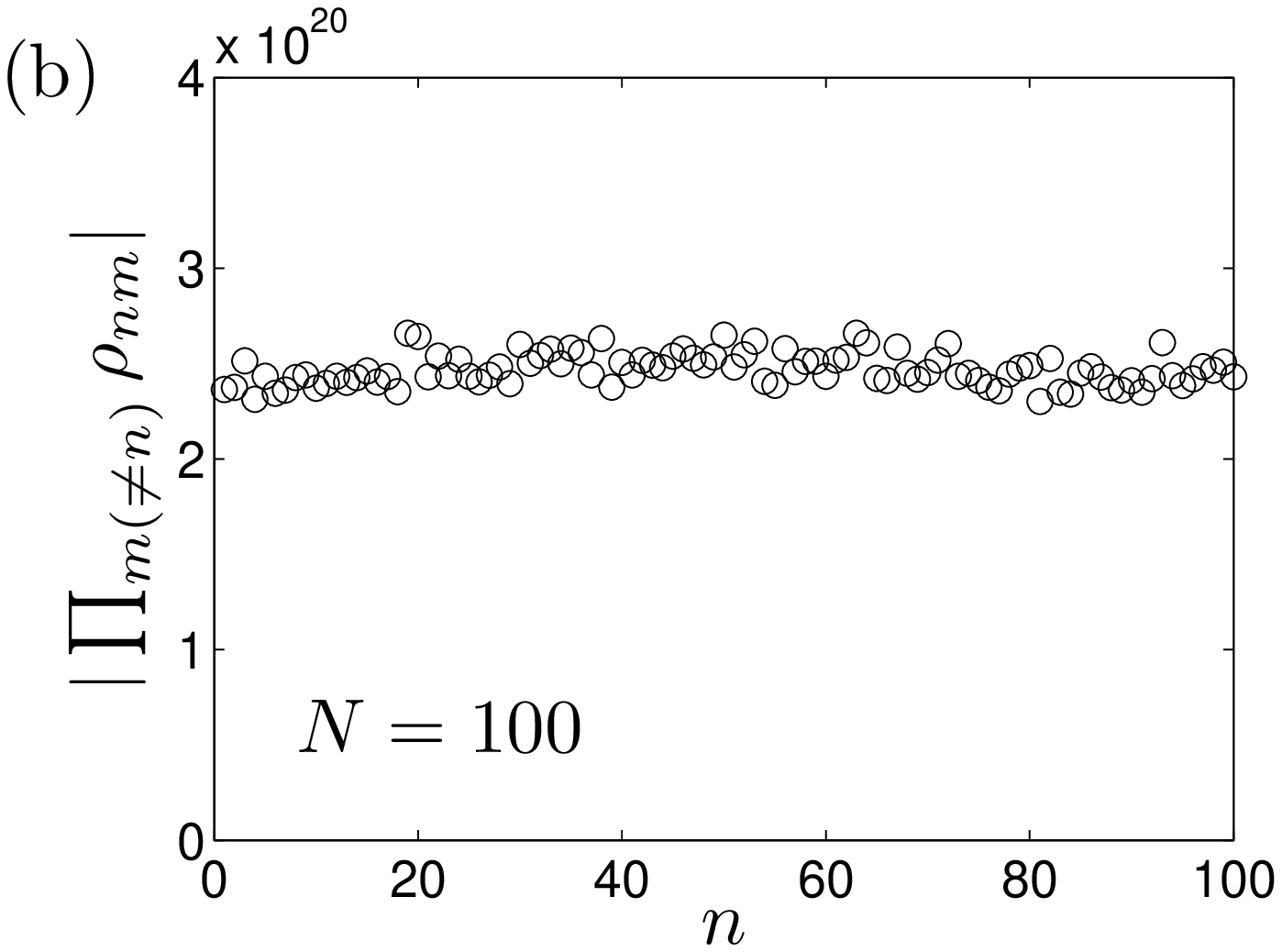}
\caption{(a) Comparison between $-\ln(|f_5(z)|)$ and $-\ln(|f_\infty(z)|)=\pi |z|^2/(2b^2)+\mathrm{constant}$. The points almost fall on a straight line with unit slope (solid line). (b) Plot of $|\prod_{m(\neq n)}\rho_{nm}|$ as a function of $n$ for $100$ spins on a sphere.}\label{figsumcheck}
\end{figure}

It has been speculated in \cite{balatsky} that the KL state is proportional to the conformal block in \eqref{wf} with $\alpha=1/2$ on an infinite square lattice. Here, we proof explicitly in the supplement \cite{supplement} for the case of a $2M\times 2M$ square lattice centered at the origin and with lattice constant $b=\sqrt{4\pi\alpha}$ that the ratio between \eqref{wf} with appropriately chosen $\chi_{p,s_p}$ and \eqref{Laughlin} with $\nu=(4\alpha)^{-1}$ is given by
\begin{equation}
\prod_{n=1}^{4M^2}|f_M(z_n)/f_\infty(z_n)|^{\alpha(1+s_n)}
\end{equation}
up to an irrelevant overall factor, where $f_M(z_n)\equiv(z_n/b)\prod_{m(\neq n)}^{4M^2}(1-z_n/z_m)^{-1}$. In particular, the two wavefunctions coincide for $M\rightarrow\infty$. In brief, we prove this result by transforming the spins into hard-core bosons by writing $s_n=2q_n-1$, $q_n\in\{0,1\}$. This allows us to express \eqref{wf} in terms of $f_M(z_n)$. We then take the limit $M\rightarrow\infty$, compute $f_\infty(z_n)$ by algebraic methods, and compare the result to \eqref{Laughlin}. We note that the correct density of particles is obtained by scaling the lattice constant rather than by changing the filling factor of the lattice. As a consequence, the lattice filling factor, which is always $1/2$, only coincides with the Landau level filling factor $\nu$ for $\alpha=1/2$. Figure~\ref{figsumcheck}(a) demonstrates that $|f_M(z)|\approx |f_\infty(z)|$ already for a $10\times10$ lattice, and thus there is a close relationship between \eqref{Laughlin} and \eqref{wf} even for small systems.

The model can be mapped from the complex plane to the unit sphere with polar angle $\theta$ and azimuthal angle $\phi$ by the stereographic projection $z=v/u$, where $u=\cos(\theta/2)e^{i\phi/2}$ and $v=\sin(\theta/2)e^{-i\phi/2}$. A small computation shows that $\rho_{ij}\equiv(v_iu_j-u_iv_j)^{-1}$ and also the wavefunction \eqref{wf} are invariant under $SU(2)$ transformations of the pair $(u,v)$. Note that $d_{ij}=2|\rho_{ij}|^{-1}$ is the shortest distance $d_{ij}=|\mathbf{n}_i-\mathbf{n}_j|$ between spin $i$ and spin $j$, where $\mathbf{n}_i\equiv(\sin(\theta_i)\cos(\phi_i),\sin(\theta_i)\sin(\phi_i),\cos(\theta_i))$ is the position vector of spin $i$. Writing $s_n=2q_n-1$ as before, we find that \eqref{wf} is proportional to the Laughlin wavefunction on the sphere $\prod_{n<m}^N\rho_{nm}^{-q_nq_m´/\nu}$ \cite{sphere} with $\nu=(4\alpha)^{-1}$ except for an extra factor of $\prod_{n\neq m}\rho_{nm}^{2\alpha q_n}$. Figure~\ref{figsumcheck}(b) shows that $|\prod_{m(\neq n)}\rho_{nm}|$ varies only little with $n$ for $N=100$, and so the correspondence between the proposed wavefunctions and the Laughlin states is again approximately valid (note that the phase of $\prod_{m(\neq n)}\rho_{nm}^{2\alpha q_n}$ can be absorbed in $\chi_{n,s_n}$). Here, and in the following, we choose the distribution of the spins on the sphere by minimizing $\sum_{i<j}d_{ij}^{-2}$ numerically.

\paragraph{Hamiltonian}

For $\alpha=1/2$, the vertex operators can be regarded as representations of spin $1/2$ fields in the $SU(2)_1$ WZW model. Using properties of null fields in this model and the Ward identity, we derive \cite{supplement} a set of positive semi-definite and Hermitian operators
\begin{multline}\label{Ham}
H_i=\frac{1}{2}\sum_{j(\neq i)}|w_{ij}|^2
-\frac{2i}{3}\sum_{j\neq k(\neq i)}\bar{w}_{ij}w_{ik}
\,\mathbf{S}_i\cdot(\mathbf{S}_j\times\mathbf{S}_k)\\
+\frac{2}{3}\sum_{j(\neq i)}|w_{ij}|^2\mathbf{S}_i\cdot \mathbf{S}_j+\frac{2}{3}\sum_{j\neq k(\neq i)}\bar{w}_{ij}w_{ik}\mathbf{S}_j\cdot \mathbf{S}_k,
\end{multline}
$i=1,\ldots,N$, which annihilate the state \eqref{wf}. In \eqref{Ham}, $w_{ij}=g(z_i)/(z_i-z_j)+h(z_i)$, where $g$ and $h$ are arbitrary functions of $z_i$, and $\mathbf{S}_i=(S_i^x,S_i^y,S_i^z)$ is the spin operator of the $i$th spin. It follows that \eqref{wf} is the ground state of $H_i$ and thus also of $H=\sum_iH_i/4+(N+1)\sum_{i,j}\mathbf{S}_i\cdot \mathbf{S}_j/6$.

\begin{figure}
\includegraphics[width=0.49\columnwidth]{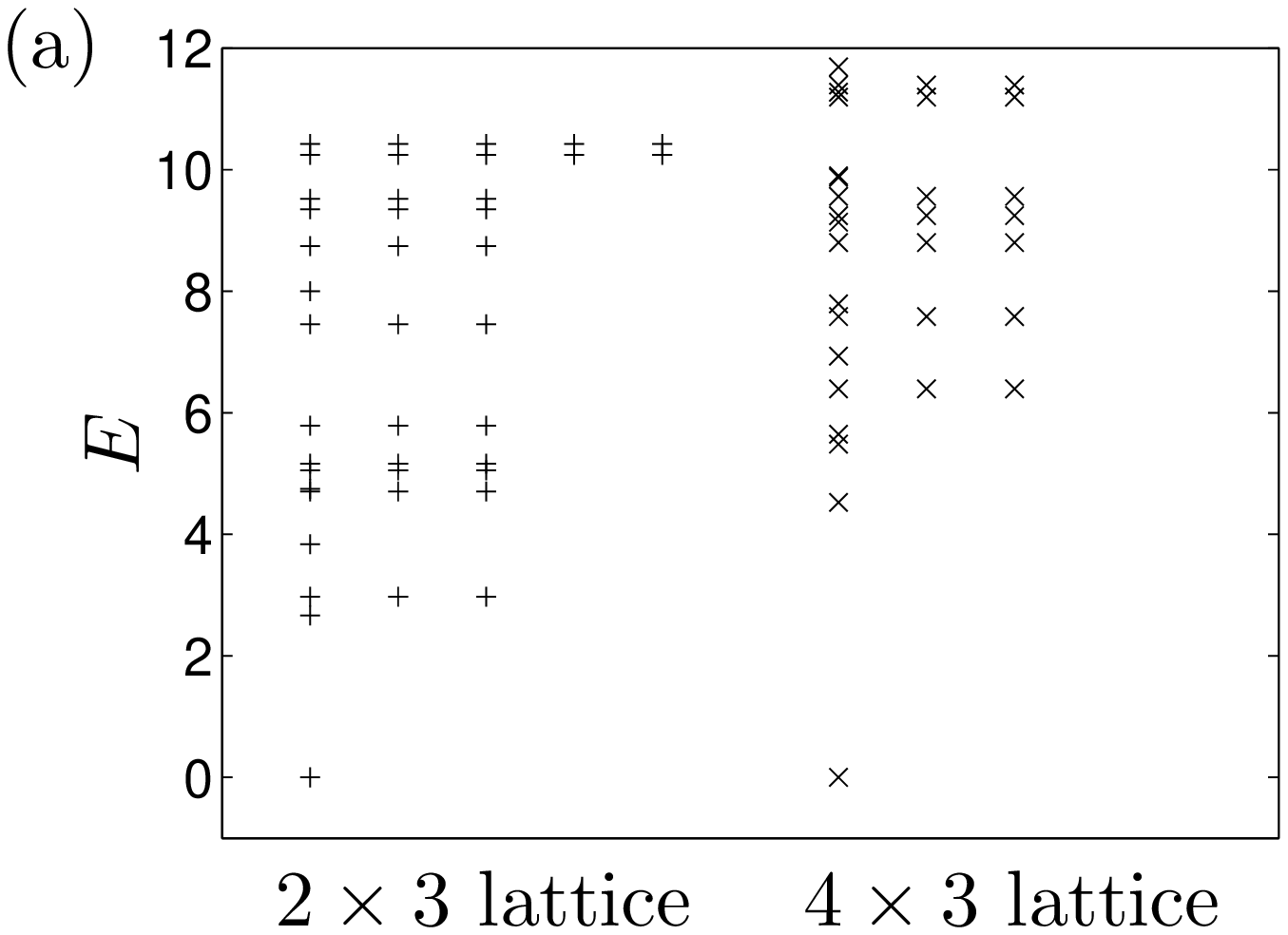}
\includegraphics[width=0.49\columnwidth]{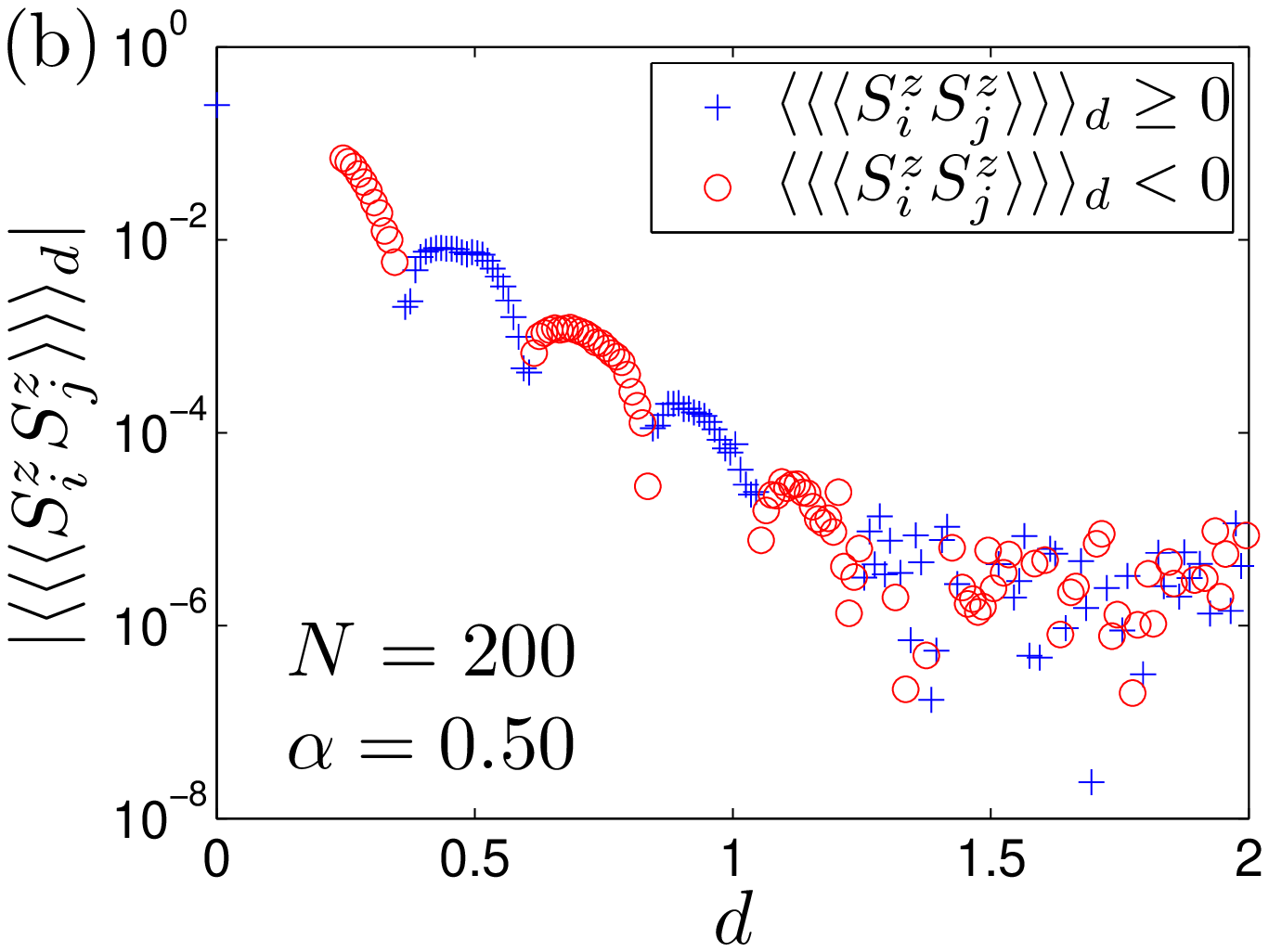}
\caption{(Color online) (a) Low-lying part of the energy spectrum for a $2\times3$ ($+$) and a $4\times3$ ($\times$) square lattice centered at the origin. (The horizontal axis shows the degeneracy.) (b) Averaged spin-spin correlation function for $200$ spins on the sphere as a function of the distance between the spins. The error bars are of order a few times $10^{-5}$, and so the results are only converged for $d\lesssim 1.2$. The symbols encode the sign of the correlations (plus for positive and circle for negative).}\label{figcor1}
\end{figure}

We note that $H$ reduces to the HS Hamiltonian \cite{H88,S88} when $z_n=\exp(2\pi i n/N)$ and $w_{ij}=2z_i/(z_i-z_j)-1$, and the construction is hence a generalization of the HS model to 2D and nonuniform distributions of the spins. In \cite{supplement}, we show that if the ground state is degenerate, then the additional ground states will not satisfy the Knizhnik-Zamolodchikov (KZ) equation \cite{KZ}. The KZ equation is derived from the Sugawara construction that builds the Virasoro generators in terms of the Kac-Moody currents. So a degeneracy would indicate theories obeying the Kac-Moody but not the Virasoro algebra. It would be surprising if such theories exist, and we therefore expect the ground state to be unique. Exact diagonalization of $H$ for small systems (see Fig.~\ref{figcor1}(a) for examples) also suggests uniqueness.

On the sphere, we can obtain a Hamiltonian, which is invariant under $SU(2)$ transformations of $(u,v)$ by choosing $H=\sum_i|u_i|^{-2}(H_i^{(1)}+H_i^{(2)})$, where $H_i^{(1)}$ ($H_i^{(2)}$) is \eqref{Ham} with $w_{ij}=1/(z_i-z_j)$ ($w_{ij}=z_i/(z_i-z_j)$). Finally, we note that a Hamiltonian for the case $\alpha=1/4$ can be constructed by summing single-particle Hamiltonians, each of which is the identity minus the sum of the projections onto the orthonormalized single-particle states.

\paragraph{Properties}

To further demonstrate the closeness between \eqref{wf} and the Laughlin states, we compute various properties of \eqref{wf} in the following. We note that all the numerical results presented below except those related to entanglement spectra are independent of $\chi_{p,s_p}$.

\paragraph{Spin-spin correlation function}

Since the systems we consider are too big for exact numerical computations, we use the Metropolis Monte Carlo algorithm to compute the spin-spin correlation function
\begin{equation}
\langle S_i^zS_j^z\rangle =\frac{\sum_{s_1,\ldots,s_N}s_is_j|\psi_{s_1,\ldots,s_N}(z_1,\ldots,z_N)|^2} {4\sum_{s_1,\ldots,s_N}|\psi_{s_1,\ldots,s_N}(z_1,\ldots,z_N)|^2}.
\end{equation}
For $\alpha=1/2$, the state is $SU(2)$ invariant and so $\langle S_i^aS_j^b\rangle=\delta_{ab}\langle S_i^zS_j^z\rangle$, $a,b=x,y,z$. Figure \ref{figcor1}(b) shows the average of the spin-spin correlation function for $200$ spins on the sphere and $\alpha=1/2$ as a function of the distance between the spins. For a given $d$, the average is taken over all spin pairs for which the distance $d_{ij}$ between the spins falls within the interval $[d-\epsilon,d+\epsilon[$, where $\epsilon=0.005$. As for the Laughlin state with $\nu=1/2$, we observe antiferromagnetic oscillations and exponential decay of the correlations.

\begin{figure}
\includegraphics[width=0.49\columnwidth]{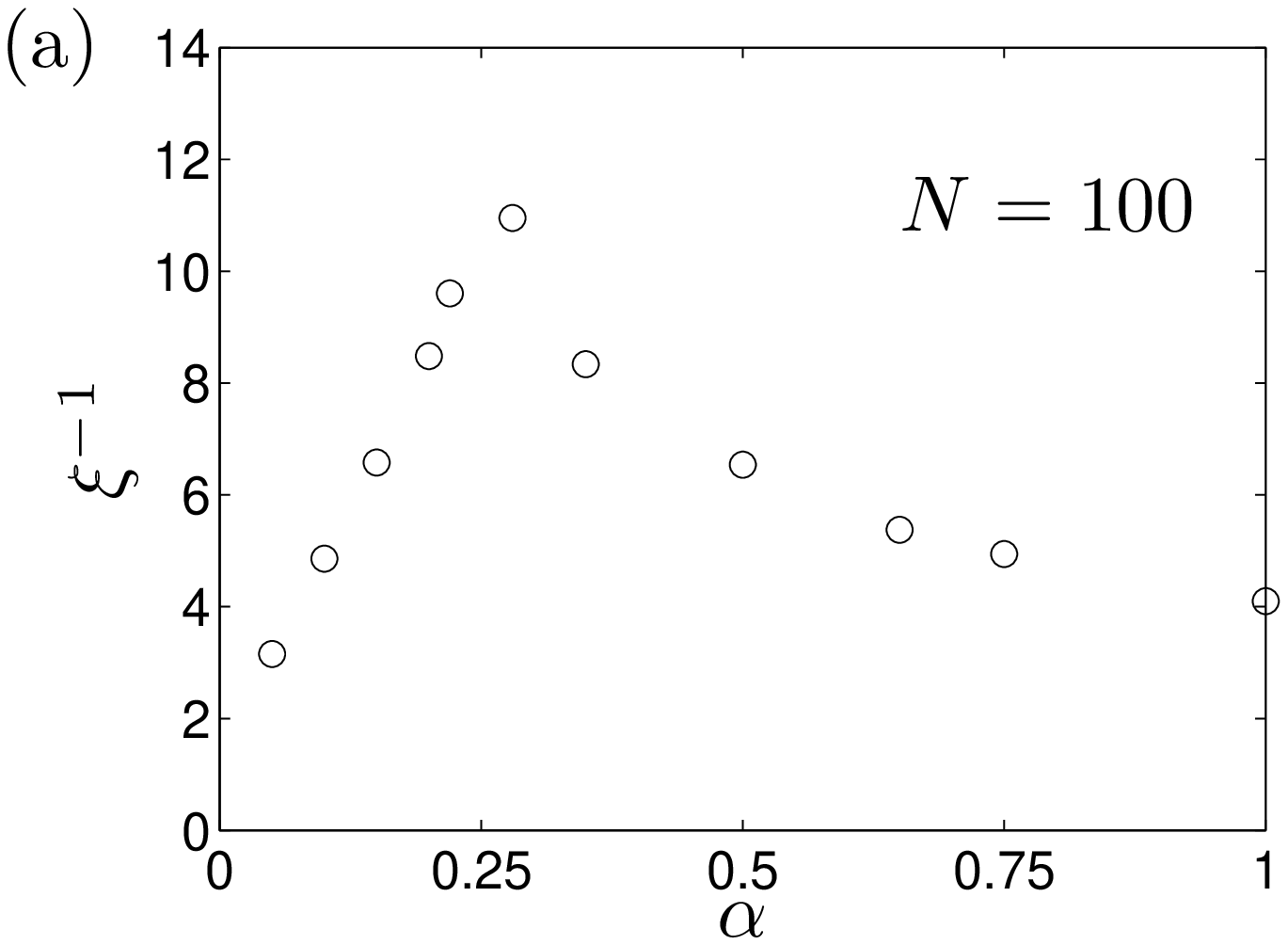}
\includegraphics[width=0.49\columnwidth]{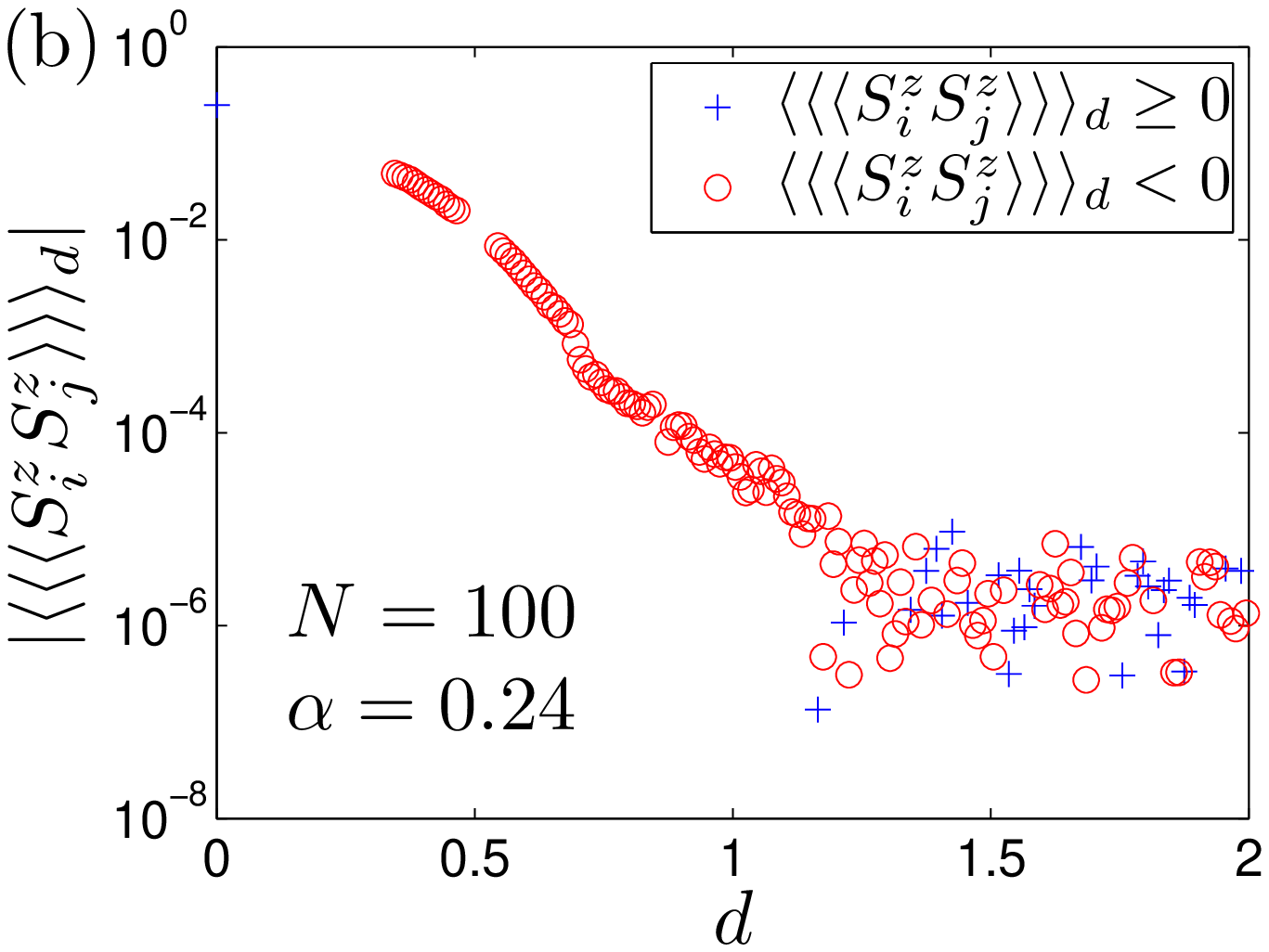}
\includegraphics[width=0.49\columnwidth]{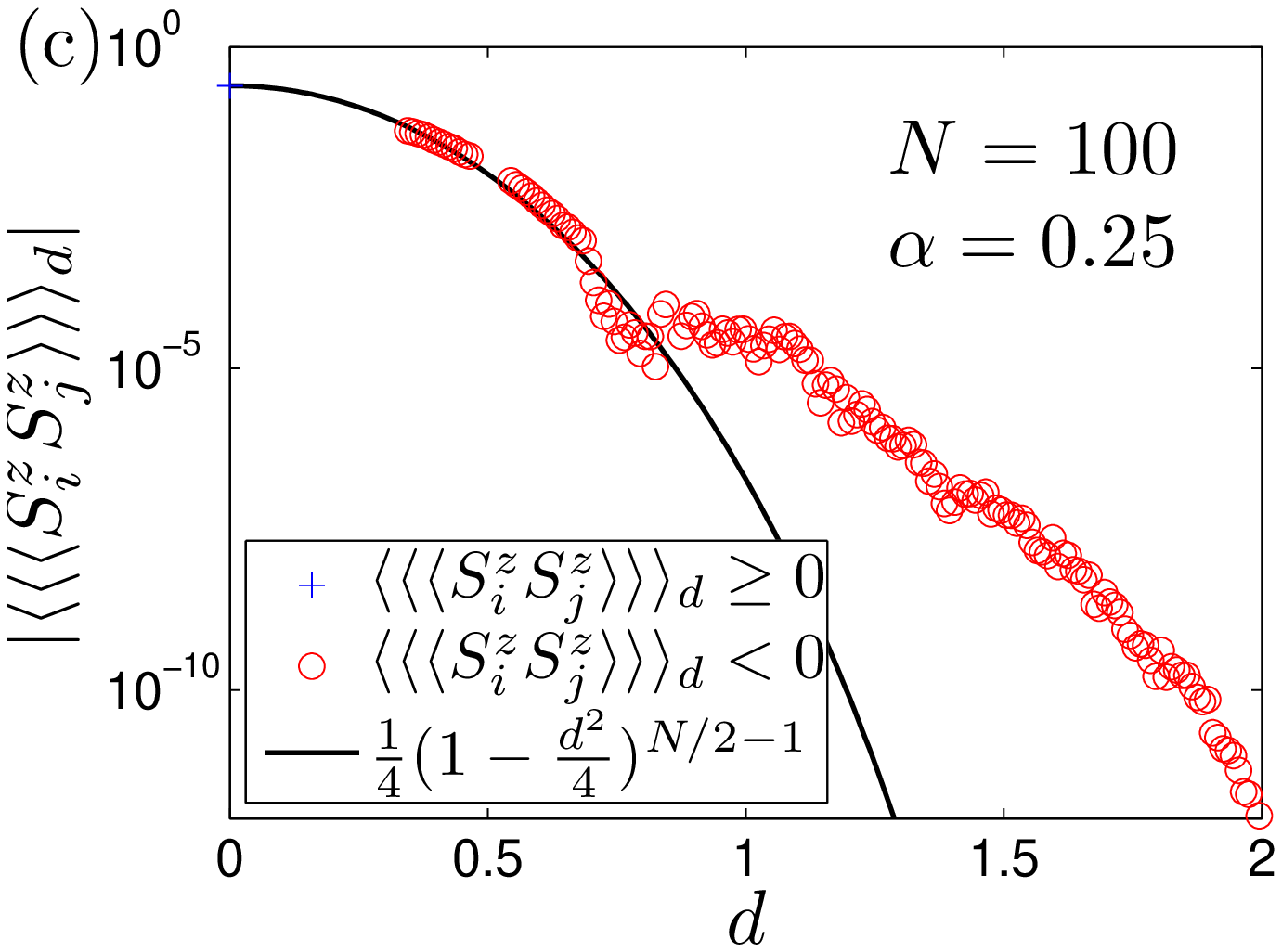}
\includegraphics[width=0.49\columnwidth]{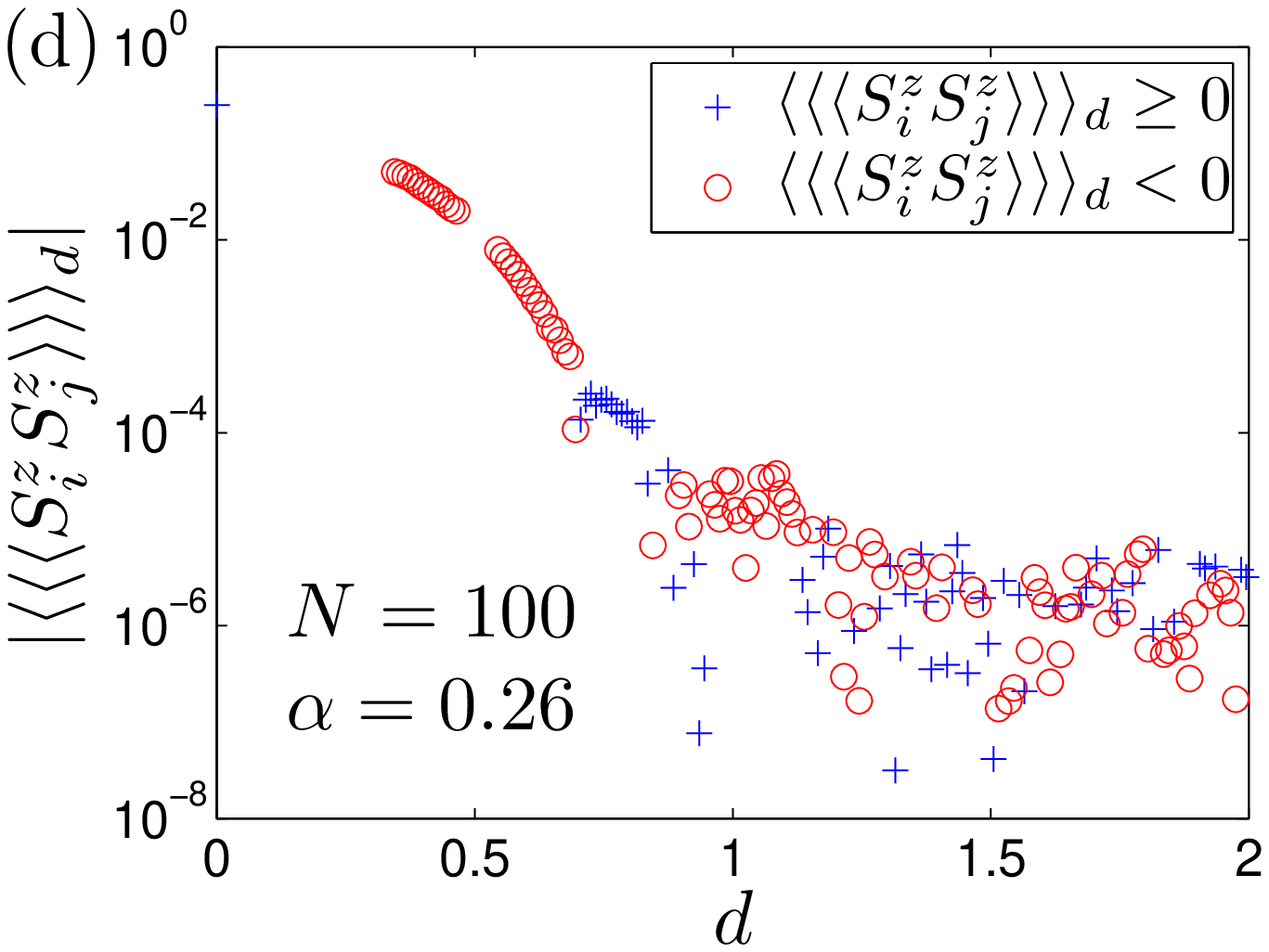}
\caption{(Color online) (a) Inverse correlation length as a function of $\alpha$ for $100$ spins on a sphere. (b-d) Averaged spin-spin correlation function as a function of the distance between the spins for (b) $\alpha=0.24$, (c) $\alpha=0.25$, and (d) $\alpha=0.26$. The error bars in (b) and (d) are of order $10^{-5}$ for all points, while the results in (c) are exact. The solid curve in (c) is minus the expression for the correlation function of the continuous $\nu=1$ Laughlin state on the sphere found in \cite{plascor3}.}\label{figcorlength}
\end{figure}

The dependence of the spin-spin correlation function on $\alpha$ is investigated in Fig.~\ref{figcorlength}. Except for $\alpha$ close to $0.25$, we find that the correlator decays approximately as $\exp(-d/\xi)$, where $d$ is the distance between the spins and $\xi$ is the correlation length. Numerical estimates of $\xi^{-1}$ are shown in Fig.~\ref{figcorlength}(a). We also find that antiferromagnetic oscillations occur above $\alpha=0.25$, but not below. This observation is consistent with the lack of oscillations in the correlation function for the $\nu=1$ Laughlin state and the presence of oscillations for filling factors below unity. It is also consistent with the conjecture that the transition occurs precisely at $\nu=1$ \cite{plascor1,plascor2}. The transition is illustrated in plots (b-d). Finally, an analytical expression for the correlation function of the continuous $\nu=1$ Laughlin state on the sphere has been found in \cite{plascor3}, and the figure shows good agreement on intermediate length scales.

\paragraph{Entanglement entropy}

The possibility of having quasiparticles with fractional statistics is a very important aspect of the Laughlin states, and it is therefore very relevant to check whether \eqref{wf} also has nontrivial topological properties. Here, we study the entanglement entropy (EE), since this allows us to extract the total quantum dimension $D$ \cite{FFN07}. More precisely, if we divide a system into two subsystems $A$ and $B$ and the system is gapped, the EE has the form $aP-\gamma+...$ \cite{Kitaev}, where $a$ is a constant, $P$ is (proportional to) the length of the boundary between $A$ and $B$, $-\gamma$ is called the topological entanglement entropy and fulfils $\gamma=\ln(D)$ \cite{Kitaev,levin}, and the ellipsis stands for terms that vanish for $P\rightarrow\infty$. For abelian systems, $D^2$ is the number of different quasiparticles one can get by fusing the fundamental quasiparticles in the system, and so $D=\sqrt{1/\nu}$ for the Laughlin states with $1/\nu\in\mathbb{N}$ \cite{FFN07}.

\begin{figure}
\includegraphics[width=0.49\columnwidth]{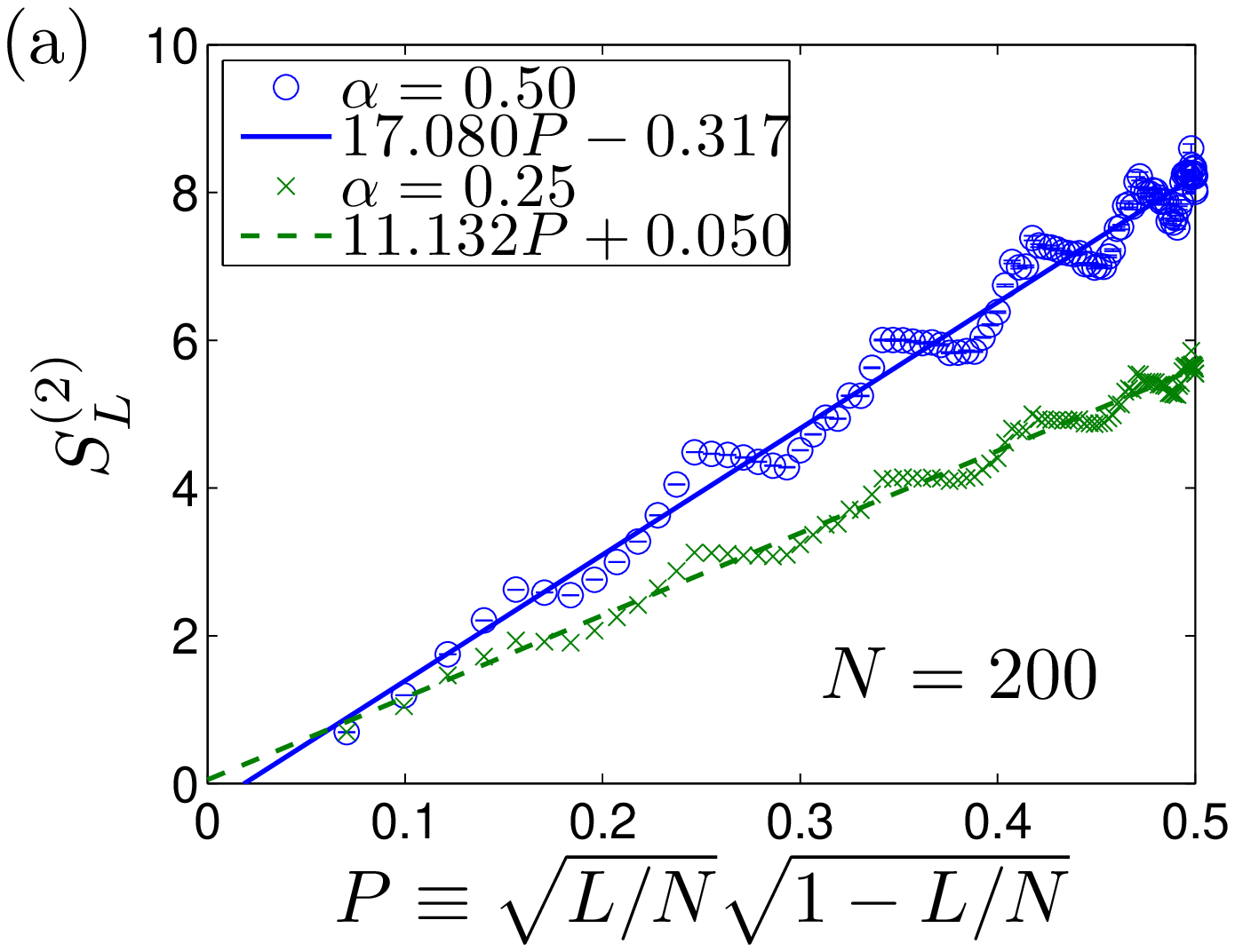}
\includegraphics[width=0.49\columnwidth]{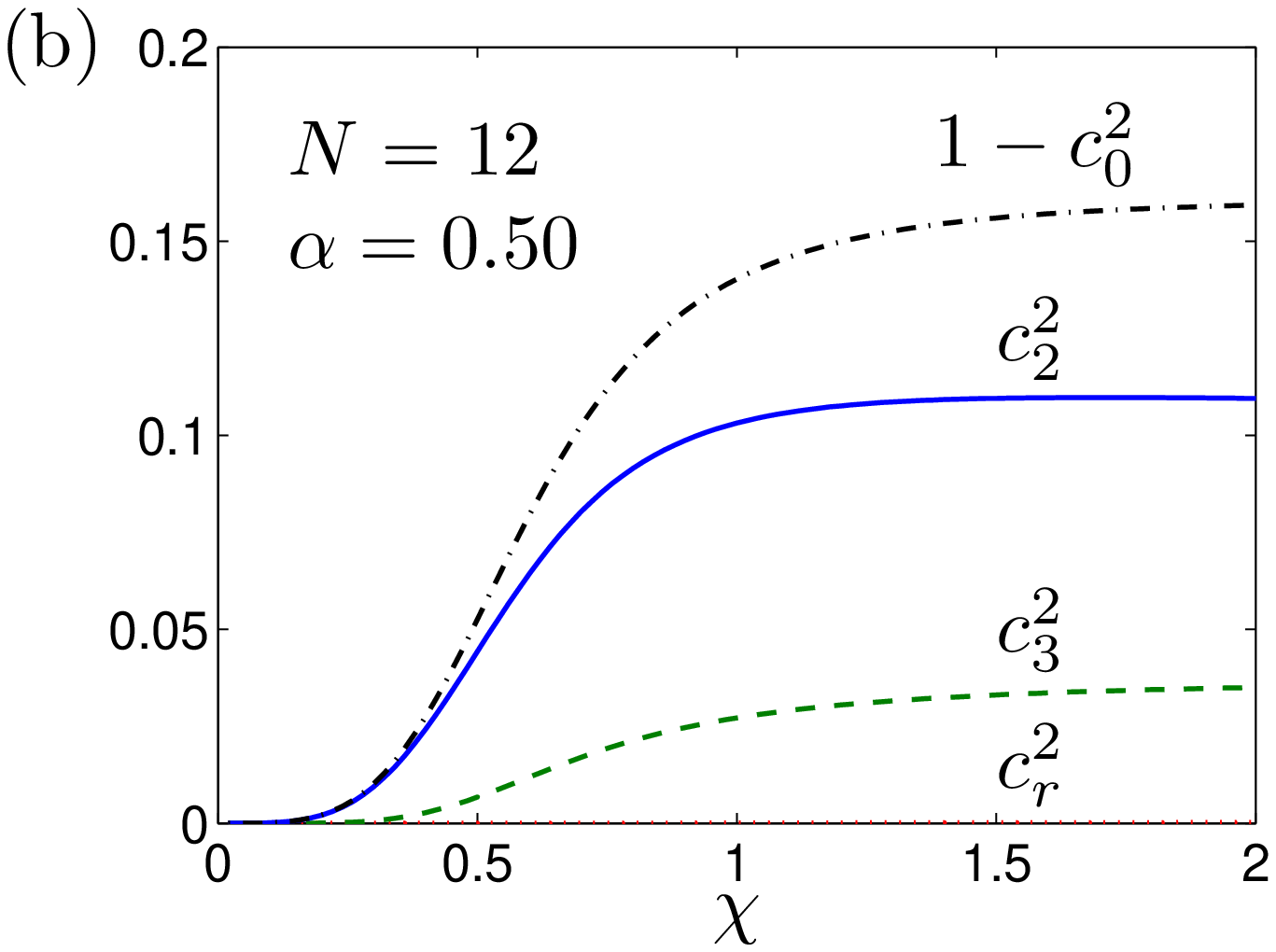}
\caption{(Color online) (a) Renyi entropy $S_L^{(2)}=-\ln(\Tr(\rho_A^2))$ for $200$ spins on a sphere for $\alpha=1/2$ (blue circles) and $\alpha=1/4$ (green crosses) when region $A$ is chosen to be the $L$ spins that are closest to the north pole. The solid and dashed lines are linear fits. (b) The coefficients in \eqref{entspec} for $N=12$.}\label{figent}
\end{figure}

The linear increase of the EE with $P$ is confirmed in Fig.~\ref{figent}(a). We use here the Renyi entropy $S_L^{(2)}=-\ln(\Tr(\rho_A^2))$, where $\rho_A$ is the reduced density operator of the spins in region $A$, because this quantity can be computed efficiently by use of Monte Carlo methods \cite{cs10,melko}. The boundary between A and B is assumed to be the circle $\theta=\theta_L$, where $\theta_L$ is defined such that the area on the sphere with $\theta<\theta_L$ relative to the complete area of the sphere equals $L/N$. The length of the boundary is then proportional to $P\equiv(L/N)^{1/2}(1-L/N)^{1/2}$, and the steps in the EE appear due to the discreteness of the positions of the spins. For the Laughlin states, there are more ways, in which the system can naturally be divided into two parts, and previous studies \cite{HaquePRL,HaquePRB} have computed entanglement entropies for both orbital partitioning, in which region $A$ involves a subset of angular momentum eigenstates, and particle partitioning, in which a particular subset of the spins comprise part $A$ of the system. We note that such complications do not appear here, since the spins are fixed at specific positions.

The topological entanglement entropy cannot be read off reliably from Fig.~\ref{figent}(a) because small errors in the linear term due to irregularities in the boundary easily become dominant. Instead, we use the method proposed in \cite{Kitaev}, which eliminates the linear term by considering a linear combination of entanglement entropies of suitably chosen regions. The results in \cite{Kitaev} are for the von Neumann entropy, but it has been shown in \cite{dong} that $\gamma=\ln(D)$ also holds for $S_L^{(2)}$ when region $A$ is a topologically trivial region. For $\alpha=1/4$, we find $\gamma=0$, and for $\alpha=1/2$, we get $-\gamma=-0.341\pm0.057$ from a Monte Carlo simulation involving $160$ spins on a sphere. Both results are consistent with the expected values ($0$ and $-\ln(2)/2\approx-0.347$, respectively) and with the results obtained for lattice models in \cite{grover}.

\paragraph{Entanglement spectrum}

It has turned out (typically from numerical studies) that the low-lying part of the entanglement spectrum, defined as the eigenvalues of $-\ln(\rho_A)$, is often related to some theory on the boundary of $A$ \cite{boundary}. Utilizing the particular structure of the wavefunctions \eqref{wf}, we can here derive such a connection analytically for the case of $\alpha=1/2$ and $N$ spins distributed uniformly on two rings in the complex plane, i.e., the spins are at the positions $\exp[2\pi(2in\pm\chi)/N]$, where $n=1,2,\ldots,N/2$. Specifically, we prove in the supplement \cite{supplement} that $\rho_A$ (the reduced density operator of the inner ring) is invariant under Yangian transformations, which means that $-\ln(\rho_A)$ is a linear combination of the invariants in the HS model. More precisely, we can write $-\ln(\rho_A)$ as a linear combination of the identity $H_0$, the two-body operator $H_2=2\sum_{n\neq m}\mathbf{S}_n\cdot\mathbf{S}_m z_nz_m/z_{nm}^2$, $z_{nm}\equiv z_n-z_m$, the three-body operator $H_3=-i\sum_{n\neq m\neq p} \mathbf{S}_n\cdot(\mathbf{S}_m\times\mathbf{S}_p) z_nz_mz_p/(z_{nm}z_{mp}z_{pn})$, and operators with higher-body interactions, which we write as $H_r$ \cite{talstra}. Considering these operators as normalized vectors $|H_i\rangle$ with inner product $\langle H_i|H_j\rangle=\Tr(H_iH_j)/(\Tr(H_i^2)\Tr(H_j^2))^{1/2}$, we can write
\begin{equation}\label{entspec}
|-\ln(\rho_A)\rangle=c_0|H_0\rangle+c_2|H_2\rangle+c_3|H_3\rangle+c_r|H_r\rangle.
\end{equation}
The coefficients are given for $N=12$ in Fig.~\ref{figent}(b). We note that the above results do not generalize to the case of more than two rings.

\paragraph{Conclusion}

We have proposed a set of spin system wavefunctions that are in many respects analogous to the Laughlin states. The proposed mapping between spin states and Laughlin states builds on CFT and demonstrates the usefulness of CFT as a tool to gain insight into many-body systems. We believe that similar mappings can be found also for other quantum Hall states. In particular, one can obtain spin analogies of the Moore-Read state by generalizing the higher level 1D spin models proposed in \cite{nsc} to 2D, which can be done straightforwardly.

The analogy between Laughlin states and spin states can be carried even further since the method proposed in \cite{MR} to incorporate quasiholes by introducing additional conformal operators can also be used for the state \eqref{wf}. As for the Laughlin states, one can interpret the square of the norm of the spin wavefunction as a particular charge distribution, and we expect this distribution to screen the quasiholes. It follows immediately from the construction that the analytic continuation properties of the wavefunctions are the same as for the Laughlin states with quasiholes. It would be interesting to investigate these ideas further.

\begin{acknowledgments}
The authors acknowledge discussions with N. Read. This work has been supported by The Carlsberg Foundation, the EU project QUEVADIS, and the grants FIS2009-11654 and QUITEMAD.
\end{acknowledgments}

\onecolumngrid
\appendix
\setcounter{equation}{0}
\newpage

\begin{center}
\textbf{Supplemental material}
\end{center}

\section{Singlet property of the wavefunction for $\alpha=1/2$}

In this section, we derive the choice of phase factors $\chi_{p,s_p}$ that ensures that the wavefunction in Eq.~(2) of the main text is a singlet for $\alpha=1/2$. The easy way to see this is to note that the vertex operators used to construct the wavefunction for $\alpha=1/2$ coincide with representations of the spin $1/2$ fields of the $SU(2)_1$ WZW model, and therefore the wavefunction is $SU(2)$ invariant if the phase factors are chosen appropriately. Since the phase factors do not depend on $z_n$, it is sufficient to determine them for $z_n=\exp(2\pi in/N)$. In this case, the wavefunction reduces to the ground state of the Haldane-Shastry model, which is a singlet, and the phase factors can therefore be read off by comparison.

One can also proof the singlet property by direct computation as follows. We define the total spin operator $\mathbf{S}=\sum_{i=1}^N\mathbf{S}_i$ with components $(S^x,S^y,S^z)$ as well as the step up $(+)$ and down $(-)$ operators $S^\pm=S^x\pm iS^y$. The $\delta_s$ in Eq.~(2) of the main paper trivially ensures that $S^z|\psi\rangle=0$, where $|\psi\rangle\equiv \sum_{s_1,\ldots,s_N}\psi_{s_1,\ldots,s_N}(z_1,\ldots,z_N)|s_1,\ldots,s_N\rangle$, and therefore the question is how one should choose the phases in order to achieve that $S^\pm|\psi\rangle=0$. Noting that
\begin{align}
S_k^x|\psi\rangle&=\frac{1}{2}\sum_{s_1,\ldots,s_N}
\psi_{s_1,\ldots,s_{k-1},-s_k,s_{k+1},\ldots,s_N}(z_1,\ldots,z_N)|s_1,\ldots,s_N\rangle,\\
S_k^y|\psi\rangle&=-\frac{i}{2}\sum_{s_1,\ldots,s_N}s_k
\psi_{s_1,\ldots,s_{k-1},-s_k,s_{k+1},\ldots,s_N}(z_1,\ldots,z_N)|s_1,\ldots,s_N\rangle,
\end{align}
we have
\begin{align}
\langle s_1,\ldots,s_N| S^-|\psi\rangle&=\frac{1}{2}\sum_{k=1}^N(1-s_k)
\psi_{s_1,\ldots,s_{k-1},-s_k,s_{k+1},\ldots,s_N}(z_1,\ldots,z_N)\\
&=\frac{1}{2}\sum_{k=1}^N(1-s_k) \delta_{\mathbf{s},k}\frac{\chi_{k,-s_k}}{\chi_{k,s_k}} (-1)^{k-1} \prod_{n(\neq k)}(z_k-z_n)^{-s_ns_k} \prod_{p=1}^N\chi_{p,s_p}\prod_{n<m}^N(z_n-z_m)^{s_ns_m/2},\label{4}
\end{align}
where $\delta_{\mathbf{s},k}=1$ for $\sum_{n(\neq k)}s_n-s_k=0$ and $\delta_{\mathbf{s},k}=0$ otherwise. Since the factor $(1-s_k)$ forces $s_k$ to be minus one, we conclude that $\langle s_1,\ldots,s_N| S^-|\psi\rangle$ is zero unless the spin configuration $\langle s_1,\ldots,s_N|$ contains $N/2-1$ spin ups and $N/2+1$ spin downs. Implicitly restricting the analysis to such configurations, we can drop the delta function in the following. The last two products in \eqref{4} can be discarded since they do not depend on $k$ and cannot be zero. The question is then whether
\begin{equation}
\sum_{k_-}F(k_-) \frac{\prod_{n_+}(z_{k_-}-z_{n_+})}
{\prod_{n_-(\neq k_-)}(z_{k_-}-z_{n_-})} \label{5}
\end{equation}
is zero, where $p_\pm$ ($p=n$ or $k$) is the subset of indices $p$ for which $s_p=\pm1$ and $F(k_-)\equiv(-1)^{k_--1}\chi_{k_-,1}/\chi_{k_-,-1}$. The dependence on $z_{n_+}$ is contained in the polynomial $\prod_{n_+}(z_{k_-}-z_{n_+})$, and since $z_{n_+}$ are arbitrary and the number of factors in $\prod_{n_+}(z_{k_-}-z_{n_+})$ is $N/2-1$, we conclude that \eqref{5} is zero in general if and only if
\begin{equation}
\sum_{k_-}F(k_-) \frac{z_{k_-}^q}
{\prod_{n_-(\neq k_-)}(z_{k_-}-z_{n_-})} \label{6}
\end{equation}
is identically zero for all $q\in\{0,1,\ldots,N/2-1\}$. Let us define the map $t$, which transforms the numbers $1,2,\ldots,N/2+1$ into the indices of the $N/2+1$ down spins, i.e.\ $\{t(1),t(2),\ldots,t(N/2+1)\}$ is the set $k_-$. Multiplying \eqref{6} by $\prod_{n_-<m_-}(z_{n_-}-z_{m_-})$, the expression turns into
\begin{equation}
D_q\equiv\sum_{p=1}^{N/2+1}F(t(p)) z_{t(p)}^q(-1)^{p-1}\prod_{\substack{n<m\\(n,m\neq p)}}^{N/2+1}(z_{t(n)}-z_{t(m)}).
\end{equation}
Comparing this to the Vandermonde determinant
\begin{equation}
V(\alpha_1,\alpha_2,\ldots,\alpha_M)\equiv\left|\begin{array}{cccc}
\alpha_1^0 & \alpha_2^0 & \cdots & \alpha_M^0\\
\alpha_1^1 & \alpha_2^1 & \cdots & \alpha_M^1\\
\vdots & \vdots & \ddots & \vdots\\
\alpha_1^{M-1} & \alpha_2^{M-1} & \cdots & \alpha_M^{M-1}
\end{array}\right|=(-1)^{M(M-1)/2}\prod_{1\leq n<m\leq M}(\alpha_n-\alpha_m),
\end{equation}
we get
\begin{align}
D_q&=(-1)^{N(N/2-1)/4}\sum_{p=1}^{N/2+1}F(t(p)) z_{t(p)}^q(-1)^{p-1}V(z_{t(1)},z_{t(2)},\ldots,z_{t(p-1)}, z_{t(p+1)},z_{t(p+2)}\ldots,z_{t(N/2+1)})\\
&=(-1)^{N(N/2-1)/4}\left|\begin{array}{cccc}
F(t(1))z_{t(1)}^q & F(t(2))z_{t(2)}^q & \cdots & F(t(\frac{N}{2}+1))z_{t(\frac{N}{2}+1)}^q\\
z_{t(1)}^0 & z_{t(2)}^0 & \cdots & z_{t(\frac{N}{2}+1)}^0\\
z_{t(1)}^1 & z_{t(2)}^1 & \cdots & z_{t(\frac{N}{2}+1)}^1\\
\vdots & \vdots & \ddots & \vdots\\
z_{t(1)}^{\frac{N}{2}-1} & z_{t(2)}^{\frac{N}{2}-1} & \cdots & z_{t(\frac{N}{2}+1)}^{\frac{N}{2}-1}
\end{array}\right|.
\end{align}
For $F(t(p))\neq0$ and general $z_{t(p)}$, this determinant is zero for all $q\in\{0,1,2,\ldots,N/2-1\}$ if and only if $F(t(p))=1$ for all $p$. The conclusion is thus that in order to achieve $\langle s_1,\ldots,s_N|S^-|\psi\rangle=0$ for all spin configurations $\langle s_1,\ldots,s_N|$, we need to choose $\chi_{k,-1}=(-1)^{k-1}\chi_{k,1}$ for $k=1,2,\ldots,N$, which means $\chi_{k,s_k}=\exp(i\frac{\pi}{2}(k-1)(s_k+1))$ up to an overall irrelevant constant factor. For this choice of $\chi_{k,s_k}$, the wavefunction transforms as $|\psi\rangle\rightarrow(-1)^{N/2}|\psi\rangle$ when all the spins are rotated by $\pi$ around the $x$-axis. The same transformation transforms $S^{\pm}$ into $S^{\mp}$, and therefore it follows immediately that also $S^+|\psi\rangle=0$.

\section{Proof that the proposed wavefunctions coincide with the Laughlin states\\ in the limit of an infinite square lattice}

We now proof that the wavefunctions in Eq.~(2) of the main paper coincide with the Laughlin states in the case of an infinite square lattice with lattice constant $b=\sqrt{4\pi\alpha}$. The first step, which is valid for general $z_n$, is to transform the spins into hard-core bosons by writing $s_n=2q_n-1$, where $q_n\in\{0,1\}$. Dropping irrelevant $q_n$-independent factors, this gives
\begin{equation}
\delta_\mathbf{s}\prod_{n<m}^N(z_n-z_m)^{\alpha s_ns_m}
\propto\delta_\mathbf{q}\prod_{n<m}^N(z_n-z_m)^{4\alpha q_nq_m}
\prod_{n=1}^N(-1)^{-2\alpha(n-1)q_n}
\prod_{n\neq m}^N(z_n-z_m)^{-2\alpha q_n},
\end{equation}
where $\delta_\mathbf{q}=1$ for $\sum_{n=1}^Nq_n=N/2$ and $\delta_\mathbf{q}=0$ otherwise. Since
\begin{align}
\delta_\mathbf{q}\prod_{n\neq m}^N(z_n-z_m)^{-2\alpha q_n}
&=\delta_\mathbf{q}\prod_{n\neq m}^N(-z_m)^{-2\alpha q_n}
\prod_{n\neq m}^N(1-z_n/z_m)^{-2\alpha q_n}\\
&=\delta_\mathbf{q}\prod_{m=1}^N(-z_m)^{-2\alpha (N/2-q_m)}
\prod_{n\neq m}^N(1-z_n/z_m)^{-2\alpha q_n},
\end{align}
we get
\begin{equation}\label{factor}
\delta_\mathbf{s}\prod_{n<m}^N(z_n-z_m)^{\alpha s_ns_m}\propto\delta_\mathbf{q}\prod_{n<m}^N(z_n-z_m)^{4\alpha q_nq_m} \prod_{n=1}^N(-1)^{-2\alpha(n-1)q_n}\prod_{n=1}^N[f_M(z_n)]^{2\alpha q_n},
\end{equation}
where $f_M(z_n)\equiv(z_n/b)\prod_{m(\neq n)}^N(1-z_n/z_m)^{-1}$ and the label $M=\sqrt{N}/2$ is introduced, because we will specialize to the case of a $2M\times 2M$ lattice centered at the origin in the following, i.e.\ $z_n=b(n_1+in_2)$, where $n=(n_1,n_2)$ and $n_1,n_2=-M+\frac{1}{2},-M+\frac{3}{2},\ldots,M-\frac{1}{2}$.

To compute $f_\infty(z)$ on this square lattice, we note that
\begin{align}
f_\infty(z+b)&=(z/b+1)\lim_{M\rightarrow\infty}\prod_{\substack{m_1+im_2\neq z/b+1\\ -M+\frac{1}{2}\leq m_1,m_2 \leq M-\frac{1}{2}}}\left(1-\frac{z/b+1}{m_1+im_2}\right)^{-1}\nonumber\\
&=(z/b+1)\lim_{M\rightarrow\infty}\prod_{\substack{m_1+im_2\neq z/b+1\\ -M+\frac{1}{2}\leq m_1,m_2 \leq M-\frac{1}{2}}}\left[\frac{m_1+im_2}{m_1-1+im_2} \left(1-\frac{z/b}{m_1-1+im_2}\right)^{-1}\right]\nonumber\\
&=\frac{z}{b}\lim_{M\rightarrow\infty}\prod_{-M+\frac{1}{2}\leq m_1,m_2 \leq M-\frac{1}{2}}\frac{m_1+im_2}{m_1-1+im_2}\prod_{\substack{m_1+im_2\neq z/b\\ -M-\frac{1}{2}\leq m_1 \leq M-\frac{3}{2}\\
-M+\frac{1}{2}\leq m_2 \leq M-\frac{1}{2}}} \left(1-\frac{z/b}{m_1+im_2}\right)^{-1}\nonumber\\
&=\lim_{M\rightarrow\infty}\prod_{-M+\frac{1}{2}\leq m \leq M-\frac{1}{2}}\frac{M-\frac{1}{2}+im}{M+\frac{1}{2}-im}\prod_{
-M+\frac{1}{2}\leq m \leq M-\frac{1}{2}} \left[\left(1+\frac{z/b}{M+\frac{1}{2}-im}\right)^{-1} \left(1-\frac{z/b}{M-\frac{1}{2}+im}\right)\right]f_\infty(z)\nonumber\\
&=\lim_{M\rightarrow\infty}\prod_{-M+\frac{1}{2}\leq m \leq M-\frac{1}{2}}\frac{M-\frac{1}{2}+im-z/b}{M+\frac{1}{2}-im+z/b}f_\infty(z)\equiv C(z)f_\infty(z).
\end{align}
From the Taylor expansion of $\ln(1+w)$, where $w$ is a complex number, we have
\begin{equation}
1+w=\exp\left(-\sum_{p=1}^\infty\frac{(-w)^p}{p}\right),
\end{equation}
and therefore
\begin{align}
C(z)&=\lim_{M\rightarrow\infty}\exp\left\{\sum_{-M+\frac{1}{2}\leq m \leq M-\frac{1}{2}}\sum_{p=1}^\infty\frac{1}{pM^p}\left[\left(im-z/b-1/2\right)^p -\left(z/b+1/2-im\right)^p\right]\right\}\nonumber\\
&=\lim_{M\rightarrow\infty}\exp\left[\sum_{q=0}^\infty\frac{2}{(2q+1)M^{2q+1}} \sum_{-M+\frac{1}{2}\leq m \leq M-\frac{1}{2}} \left(im-z/b-1/2\right)^{2q+1} \right].\label{C}
\end{align}
Consider
\begin{equation}\label{ex}
\left(im-z/b-1/2\right)^{2q+1}
=(im)^{2q+1}-(2q+1)(im)^{2q}\left(z/b+1/2\right)+O(m^{2q-1}).
\end{equation}
The sum of $m^{2q+1}$ is zero, since $m^{2q+1}$ is an odd function of $m$, and so only the term proportional to $m^{2q}$ contributes in the limit $M\rightarrow\infty$.
Combining \eqref{C} and \eqref{ex} and using the relations
\begin{equation}
\sum_{m=\frac{1}{2}}^{M-\frac{1}{2}}m^{2q}=\frac{M^{2q+1}}{2q+1}+O(M^{2q})\qquad \textrm{and}\qquad \sum_{q=0}^\infty\frac{(-1)^q}{2q+1}=\frac{\pi}{4},
\end{equation}
we find
\begin{equation}
C(z)=\lim_{M\rightarrow\infty}\exp\left[-\sum_{q=0}^\infty\frac{(-1)^q}{M^{2q+1}} \sum_{-M+\frac{1}{2}\leq m \leq M-\frac{1}{2}} m^{2q}\left(2z/b+1\right) \right]=\exp\left[-\frac{\pi}{2}(2z/b+1)\right].
\end{equation}
From this and a similar computation, we conclude
\begin{equation}
f_\infty(z+b)=\exp\left[-\frac{\pi}{2}(2z/b+1)\right]f_\infty(z)
\qquad\textrm{and}\qquad
f_\infty(z+ib)=\exp\left[-\frac{\pi}{2}(-2iz/b+1)\right]f_\infty(z).
\end{equation}
Solving these equations (remembering that $\rea(z)/b$ and $\ima(z)/b$ are half integers) leads to
\begin{equation} f_\infty(z)\propto\exp\{-i\pi[\rea(z)/b-1]\ima(z)/b\}\exp[-\pi|z|^2/(2b^2)]. \end{equation}
Inserting this into \eqref{factor} and comparing to the Laughlin states in Eq.~(1) of the main text, we conclude that the wavefunction in Eq.~(2) of the main text is proportional to the Laughlin state with filling factor $\nu=(4\alpha)^{-1}$ and gauge factor $\prod_{n=1}^N(\chi_{n,2q_n-1}(-1)^{-2\alpha(n-1)q_n}\exp\{-2\alpha q_n i\pi[\rea(z_n)/b-1]\ima(z_n)/b\})$ for $M\rightarrow\infty$. We note that the difference between this result for $\alpha=1/2$ and the gauge factor in \cite{laughlinHam} is due to the fact that the lattice points are at half-integer rather than integer positions.

\section{Derivation of the Hamiltonian}

The main idea in the derivation of the parent Hamiltonian for the spin wavefunction with $\alpha=1/2$, which we presented recently in a more general setting in \cite{nsc}, is to use null fields in the $SU(2)_1$ WZW model. A null field $\chi_{s_i}(z_i)$ has the property that
\begin{equation}\label{nullfield}
\langle \phi_{s_1}(z_1)\cdots\phi_{s_{i-1}}(z_{i-1})
\chi_{s_i}(z_i)\phi_{s_{i+1}}(z_{i+1})\ldots\phi_{s_N}(z_N)\rangle=0.
\end{equation}
Here, we specifically consider the null fields \cite{nsc}
\begin{equation}
[\chi_{s_i}(z_i)]_a=\frac{2}{3}\sum_b\sum_{s'_i} \left(\delta_{ab}\delta_{s_is'_i}-i\sum_c\vep_{abc}(S_i^c)_{s_is'_i}\right)(J_{-1}^b\phi_{s'_i})(z_i),
\end{equation}
where $a,b,c$ can take the values $x,y,z$, $\vep_{abc}$ is the Levi-Civita symbol, $(S_i^c)_{s_is'_i}$ are the matrix elements of the operator $S_i^c$, and $J_{-1}^b$ is one of the modes of the chiral currents in the $SU(2)_1$ WZW model. The next step is to use the Ward identity
\begin{multline}
\langle \phi_{s_1}(z_1) \dots  \phi_{s_{i-1}}(z_{i-1})(J_{-1}^b \phi_{s'_i})(z_i)\phi_{s_{i+1}}(z_{i+1})\dots\phi_{s_N}(z_N)\rangle\\
=\sum_{j (\neq i)}^N \sum_{s'_j}\frac{(S^b_j)_{s_js'_j}}{z_j-z_i}
\langle \phi_{s_1}(z_1) \ldots \phi_{s'_i}(z_i) \ldots \phi_{s'_j}(z_j) \dots \phi_{s_N}(z_N) \rangle
\end{multline}
to rewrite \eqref{nullfield} into an expression containing the chiral correlator, which defines the spin wavefunction. This gives
\begin{equation}\label{res1}
\frac{2}{3}\sum_{j (\neq i)}^N \frac{1}{z_j-z_i} \left(S^a_j-i\sum_{b,c}\vep_{abc}S_i^cS^b_j\right)
|\psi\rangle=0.
\end{equation}
Due to the singlet property of the wavefunction, we also have
\begin{equation}\label{res2}
\frac{2}{3}\sum_{j (\neq i)}^N \left(S^a_j-i\sum_{b,c}\vep_{abc}S_i^cS^b_j\right)
|\psi\rangle=
\frac{2}{3}\left(-S^a_i+i\sum_{b,c}\vep_{abc}S_i^cS^b_i\right)
|\psi\rangle=0,
\end{equation}
where we have used the relations
\begin{equation}\label{relations}
S_i^aS_i^b=\frac{1}{4}\delta_{ab}+\frac{i}{2}\sum_c\vep_{abc}S_i^c
\qquad\textrm{and}\qquad
\sum_a\vep_{abc}\vep_{ab'c'}=\delta_{bb'}\delta_{cc'}-\delta_{bc'}\delta_{cb'}
\end{equation}
in the last step. Making an arbitrary linear combination $-g(z_i)\times(\textrm{Eq.~}\eqref{res1})+h(z_i)\times(\textrm{Eq.~}\eqref{res2})$ of the two results, where $g(z_i)$ and $h(z_i)$ are functions of $z_i$, we therefore conclude that the operator
\begin{equation}\label{ci}
\mathbf{C}_i = \frac{2}{3} \sum_{j (\neq i)}^N  w_{i j} \left( \mathbf{S}_{j} + i \mathbf{S}_i \times \mathbf{S}_{j} \right)
\end{equation}
with $w_{ij}\equiv g(z_i)/(z_i-z_j)+h(z_i)$ annihilates the wavefunction, i.e., $\mathbf{C}_i|\psi\rangle=0$. A hermitian and positive semi-definite Hamiltonian can then be defined as
\begin{equation}\label{hi}
H_i=\mathbf{C}_i^\dag \cdot \mathbf{C}_i.
\end{equation}
The result in Eq.~(4) of the main text now follows by combining \eqref{ci} and \eqref{hi} and applying \eqref{relations}. Note that the Hamiltonian in Eq.~(4) of the main text is the generalization to 2D of the 1D Hamiltonian in Eq.~(69) of \cite{nsc}.

\section{Uniqueness of the state that is both a ground state and a solution to the Knizhnik-Zamolodchikov equation}

In order for a state to be a ground state of the Hamiltonian $H$, it must be annihilated individually by all the operators $\mathcal{C}_i^a$ and $\sum_{i}S_i^a$. This means that the state must fulfil \eqref{res1} and be a singlet. The wavefunction in Eq.~(2) of the main text with $\alpha=1/2$ furthermore satisfies the Knizhnik-Zamolodchikov equation
\begin{equation}\label{KZ}
\frac{\partial}{\partial z_i}|\psi\rangle=\frac{2}{3}\sum_{j(\neq i)}\frac{\mathbf{S}_i\cdot\mathbf{S}_j}{z_i-z_j}|\psi\rangle.
\end{equation}
We now show that there is only one state, which fulfils both \eqref{res1} and \eqref{KZ} and is a singlet. To do so, we first multiply \eqref{KZ} by $S^a_i$ and use \eqref{relations} to obtain
\begin{equation}
S^a_i\frac{\partial}{\partial z_i}|\psi\rangle=\frac{1}{3}\sum_{j(\neq i)}\left(\frac{S^a_j}{2(z_i-z_j)} -\sum_{b,c}\frac{i\vep_{abc}S_i^bS^c_j}{z_i-z_j}\right)|\psi\rangle.
\end{equation}
Using \eqref{res1}, this simplifies to
\begin{equation}
S^a_i\frac{\partial}{\partial z_i}|\psi\rangle=\sum_{j(\neq i)}\frac{S^a_j}{2(z_i-z_j)}|\psi\rangle.
\end{equation}
Taking $a=z$ and utilizing $s_i^2=1$, it follows that
\begin{equation}
\frac{\partial}{\partial z_i}\psi_{s_1,\ldots,s_N}(z_1,\ldots,z_N)=\sum_{j(\neq i)}\frac{s_is_j}{2(z_i-z_j)}\psi_{s_1,\ldots,s_N}(z_1,\ldots,z_N),
\end{equation}
but this set of equations has the unique solution
\begin{equation}
\psi_{s_1,\ldots,s_N}(z_1,\ldots,z_N)=c(s_1,\ldots,s_N)\prod_{i<j}(z_i-z_j)^{s_is_j/2},
\end{equation}
where $c$ is a function, which does not depend on $z_n$. The requirement that $|\psi\rangle$ is a singlet uniquely determines $c(s_1,\ldots,s_N)$ up to an irrelevant overall constant factor. Therefore the state is unique and given by Eq.~(2) in the main text with $\alpha=1/2$.

\section{Invariance of $\rho_A$ under Yangian transformations}

From Eq.~\eqref{ci}, we know that the operator $\mathbf{Q}_i=\mathbf{Q}^{(1)}_i+\mathbf{Q}^{(2)}_i$, where
\begin{equation}
\mathbf{Q}^{(1)}_i\equiv\sum_{j(\neq i)}w_{ij}\; \mathbf{S}_j,
\qquad\mathbf{Q}^{(2)}_i\equiv i\sum_{j(\neq i)}w_{ij}\; \mathbf{S}_i\times \mathbf{S}_j,
\end{equation}
annihilates the spin wavefunction when $\alpha=1/2$. We consider here the case $w_{ij}=2z_i/(z_i-z_j)-1=(z_i+z_j)/(z_i-z_j)$ and take the $N$ spins to be uniformly distributed on $R$ rings, i.e., the complex coordinates $z_i$ are given by
\begin{equation}
z_{rn}=u_r\zeta^n,\qquad \zeta\equiv\exp(2\pi i/P),
\qquad r=1,2,\ldots,R,\qquad n=1,2,\ldots,P,
\end{equation}
where $u_r$ is the radius of the $r$th ring and $P$ is the number of spins in one ring. Note that
\begin{equation}
\sum_{n=1}^Pw_{rn,r'n'}
=\sum_{n=1}^P\frac{u_r\zeta^n+u_{r'}\zeta^{n'}}{u_r\zeta^n-u_{r'}\zeta^{n'}}
=\sum_{n=1}^P\frac{u_r\zeta^{n-n'}+u_{r'}}{u_r\zeta^{n-n'}-u_{r'}}
=\sum_{n=1}^P\frac{u_r\zeta^n+u_{r'}}{u_r\zeta^n-u_{r'}}
\equiv F_{rr'},\quad r\neq r',
\end{equation}
and
\begin{equation}
\sum_{n(\neq n')}w_{rn,rn'}
=\sum_{n(\neq n')}\frac{\zeta^{n-n'}+1}{\zeta^{n-n'}-1}
=\sum_{n=1}^{P-1}\frac{\zeta^n}{\zeta^n-1}+ \sum_{n=1}^{P-1}\frac{1}{\zeta^n-1}
=-\sum_{n=1}^{P-1}\frac{1}{\zeta^{-n}-1}+ \sum_{n=1}^{P-1}\frac{1}{\zeta^n-1}=0.
\end{equation}
Defining $F_{rr}=0$, we can write
\begin{equation}
\sum_{n=1}^P\mathbf{Q}^{(1)}_{rn}=\sum_{n\neq n'}w_{rn,rn'}\; \mathbf{S}_{rn'}+\sum_{r'(\neq r)} \sum_{n'=1}^P\sum_{n=1}^Pw_{rn,r'n'}\; \mathbf{S}_{r'n'}
=\sum_{r'=1}^R F_{rr'} \sum_{n'=1}^P\mathbf{S}_{r'n'}
\end{equation}
and hence
\begin{equation}
\sum_{r=1}^R\sum_{r''=1}^R\sum_{n=1}^P(F^{-1})_{rr''}\mathbf{Q}^{(1)}_{r''n}
=\sum_{r=1}^R\sum_{r''=1}^R(F^{-1})_{rr''}\sum_{r'=1}^R F_{r''r'} \sum_{n'=1}^P\mathbf{S}_{r'n'}
=\sum_{r'=1}^R\sum_{n'=1}^P\mathbf{S}_{r'n'}=\mathbf{S}.
\end{equation}
The latter operator is the total spin operator. Since $|\psi\rangle$ is a singlet by construction, we have $\mathbf{S}|\psi\rangle=0$. It follows from this and $\mathbf{Q}_{rn}|\psi\rangle=0$ that $|\psi\rangle$ is also annihilated by the operator
\begin{equation}\label{Q2}
\sum_{r=1}^R\sum_{r'=1}^R\sum_{n=1}^P(F^{-1})_{rr'}\mathbf{Q}^{(2)}_{r'n}
=\sum_{r=1}^R\sum_{r'=1}^R(F^{-1})_{rr'}\left(2\mathbf{\Lambda}_{r'}+i\sum_{r''(\neq r')}\sum_{n=1}^P\sum_{n'=1}^Pw_{r'n,r''n'}\; \mathbf{S}_{r'n}\times \mathbf{S}_{r''n'}\right),
\end{equation}
where
\begin{equation}
\mathbf{\Lambda}_r=\frac{i}{2}\sum_{n'\neq n}w_{rn,rn'}\; \mathbf{S}_{rn}\times \mathbf{S}_{rn'}
\end{equation}
is the Yangian of the $r$th ring. For the special case $R=2$, there is only one term in the sum over $r''$ in \eqref{Q2}, namely $r''=r$. Since $F^{-1}$, $w$, and the cross product are all antisymmetric under exchange of the indices, the last term on the right hand side of \eqref{Q2} is zero in this case, and we get the simplified expression
\begin{equation}
\sum_{r=1}^2\sum_{r'=1}^2\sum_{n=1}^P(F^{-1})_{rr'}\mathbf{Q}^{(2)}_{r'n}
=2(F^{-1})_{12}\mathbf{\Lambda}_2+2(F^{-1})_{21}\mathbf{\Lambda}_1
=2(F^{-1})_{12}(\mathbf{\Lambda}_2-\mathbf{\Lambda}_1).
\end{equation}
We hence conclude that $(\mathbf{\Lambda}_2-\mathbf{\Lambda}_1)|\psi\rangle=0$, and therefore $\exp(i(\mathbf{\Lambda}_2-\mathbf{\Lambda}_1)t) |\psi\rangle=|\psi\rangle$, where $t$ is some real parameter. From this the invariance of $\rho_A$ (the reduced density operator of ring 1) under Yangian transformations follows:
\begin{equation}
\rho_A\equiv\Tr_B(|\psi\rangle\langle\psi|) =\Tr_B\left[e^{i(\mathbf{\Lambda}_2-\mathbf{\Lambda}_1)t} |\psi\rangle\langle\psi|
e^{-i(\mathbf{\Lambda}_2-\mathbf{\Lambda}_1)t}\right]
=e^{-i\mathbf{\Lambda}_1t}\Tr_B\left[e^{i\mathbf{\Lambda}_2t} |\psi\rangle\langle\psi|
e^{-i\mathbf{\Lambda}_2t}\right]e^{i\mathbf{\Lambda}_1t}
=e^{-i\mathbf{\Lambda}_1t}\rho_Ae^{i\mathbf{\Lambda}_1t}.
\end{equation}
This result does not generalize to the case of more than two rings since there are more terms in the sum over $r''$ in \eqref{Q2} in this case.

\end{document}